\documentclass[twocolumn,showpacs,preprintnumbers,amsmath,amssymb]{revtex4}

\usepackage{graphicx}
\usepackage{dcolumn}
\usepackage{bm}

\newcommand{\bq}{\begin{equation}}
\newcommand{\eq}{\end{equation}}
\newcommand{\bqa}{\begin{eqnarray}}
\newcommand{\eqa}{\end{eqnarray}}
\newcommand{\nn}{\nonumber \\}

\def\be     {\begin{equation}}
\def\ee     {\end{equation}}
\def\bea        {\begin{eqnarray}}
\def\eea        {\end{eqnarray}}
\def\bnn    {\begin{eqnarray*}}
\def\enn    {\end{eqnarray*}}

\begin{document}

\title{Deconfined quantum criticality of the $O(3)$ nonlinear
$\sigma$ model in two spacial dimensions: A renormalization group
study}
\author{Ki-Seok Kim}
\affiliation{Korea Institute for Advanced Study, Seoul 130-012,
Korea}
\date{\today}

\begin{abstract}
We investigate the quantum phase transition of the $O(3)$
nonlinear $\sigma$ model {\it without Berry phase} in two spacial
dimensions. Utilizing the $CP^{1}$ representation of the nonlinear
$\sigma$ model, we obtain an effective action in terms of bosonic
spinons interacting via compact U(1) gauge fields. Based on the
effective field theory, we find that the bosonic spinons are
deconfined to emerge at the quantum critical point of the
nonlinear $\sigma$ model. It is emphasized that the deconfinement
of spinons is realized {\it in the absence of Berry phase}. This
is in contrast to the previous study of Senthil et al. [Science
{\bf 303}, 1490 (2004)], where the Berry phase plays a crucial
role, resulting in the deconfinement of spinons. It is the reason
why the deconfinement is obtained even in the absence of the Berry
phase effect that the quantum critical point is described by the
XY ("neutral") fixed point, not the IXY ("charged") fixed point.
The IXY fixed point is shown to be unstable against instanton
excitations and the instanton excitations are proliferated. {\it
At the IXY fixed point it is the Berry phase effect that
suppresses the instanton excitations, causing the deconfinement of
spinons}. {\it On the other hand, the XY fixed point is found to
be stable against instanton excitations because an effective
internal charge is zero at the neutral XY fixed point}. As a
result the deconfinement of spinons occurs at the quantum critical
point of the $O(3)$ nonlinear $\sigma$ model in two dimensions.
\end{abstract}


\maketitle

\section{Motivation and Summary}

Nature of quantum criticality is one of the central interests in
modern condensed matter physics. Especially, {\it deconfined
quantum criticality} has been proposed in various strongly
correlated electron systems such as low dimensional quantum
antiferromagnets\cite{Laughlin,Berry_phase,Japan_group,KT,Large_N_limit,Kim_disorder,Kim_dissipation,K_QED}
and Kondo
systems\cite{Senthil_Kondo,Coleman_Kondo,Pepin_Kondo,Kim_Kondo}.
In the present paper we investigate one deconfined quantum
criticality based on the O(3) nonlinear $\sigma$ model describing
a quantum phase transition from antiferromagnetism to quantum
disordered paramagnetism on two dimensional square lattices. This
phase transition has been originally analyzed by Bernevig et
al.\cite{Laughlin}. In the study the authors got to the conclusion
that although the appropriate "off-critical" elementary degrees of
freedom are given by either spin $1$ excitons (gapped paramagnons)
in the quantum disordered paramagnetism and spin $1$
antiferromagnons in the antiferromagnetism, at the quantum
critical point such excitations should {\it break up} into more
elementary spin $1/2$ excitations usually called {\it
spinons}\cite{Laughlin}. Thus, spinons emerge as true, deconfined,
elementary excitations right at the quantum critical point. This
is the precise meaning of the deconfined quantum criticality in
the context of quantum antiferromagnetism. In Fig. 1 schematic
phase diagram and proposed elementary excitations in the O(3)
nonlinear $\sigma$ model are shown.

This was challenged by Senthil et al.\cite{Berry_phase}. They
claimed that since the phase transition in Ref. \cite{Laughlin} is
supposed to fall into Landau-Ginzburg-Wilson ($LGW$) paradigm, the
spectrum at the quantum critical point should be fully
understandable only in terms of spin $1$ bosonic degrees of
freedom\cite{Berry_phase}. Senthil et al. proposed, as a possible
candidate for a deconfined quantum critical point, a direct
quantum phase transition between a Neel antiferromagnet and a
valance bond solid ($VBS$) state. In particular, in the Neel state
one gets spinon condensation. In the paramagnetic phase instanton
excitations (tunnelling events between energetically degenerate
but topologically inequivalent vacua of the U(1) gauge field in
the $CP^1$ representation of the O(3) nonlinear $\sigma$ model)
should possibly arise, whose condensation does not allow spinon
deconfinement. However, Senthil et al. argued that this is not the
case at the quantum critical point, {\it where a Berry phase term
makes instantons irrelevant} and accordingly, makes it possible to
achieve spinon deconfinement\cite{Berry_phase}. Apparently, this
would prove that it is not possible to get spinon deconfinement
{\it without Berry phase}, which would invalidate the results of
Bernevig et al..

In the present paper we show that such a contradiction does not
exist. We focus our attention to the $CP^1$ representation of the
O(3) nonlinear $\sigma$ model {\it without Berry phase} (that is,
the system studied by Bernevig et al.), which leads to {\it the
two flavor Abelian Higgs model}. In such a model the basic degrees
of freedom are provided by a complex doublet of bosonic spinon
fields, plus a compact U(1) gauge field giving long range
interactions among spinons. Using a renormalization group ($RG$)
analysis, we investigate the quantum critical point of the two
flavor Abelian Higgs model. To perform an $RG$ analysis, we move
to the dual representation of the $CP^1$ action, in which the
basic fields are the vortex fields representing spin $1/2$
merons\cite{Berry_phase}. In the language of meron fields the
phase where the meron fields have zero expectation value is
associated with the Neel state, while the phase in which the meron
fields take a nonzero expectation value (vortex condensation)
corresponds to a featureless quantum disordered paramagnetic phase
(here, not the $VBS$ owing to the absence of Berry phase). In both
phases processes in which an instanton is created with an attached
vortex creation (annihilation) operator are relevant. This forbids
spinon deconfinement in either off-critical phase.

To analyze the quantum critical point of the system, we first
resort to an effective low energy action in Eq. (9), where only
phase fluctuations of the vortex fields are allowed and the
instanton term is explicitly included. The parameters of such an
action are the stiffness parameter of the vortex phase, $\kappa$,
the instanton fugacity, $y_m$, and the phase stiffness of the dual
Higgs field, $\rho$. An $RG$ analysis permits us to write down the
scaling equations in Eq. (12). When specified to the particular
case $D = 3$ (that is, a planar model at zero temperature), such
equations exhibit two quantum critical points. The former one is
at $\kappa^* = 0$, $y^*_{m} = 0$ and $\rho^* = 0$. {\it Such a
critical point, dubbed inverted XY (IXY) fixed point or "charged"
XY fixed point, is identified with the quantum phase transition
studied by Senthil et al.}\cite{Berry_phase}. The IXY fixed point
is shown to be unstable against instanton excitations ($y_{m}
\not= 0$) and the instanton excitations are proliferated. Since
condensation of vortices or instantons does not allow spinon
unbinding, this is consistent with the conclusion of Senthil and
coworkers, that is, with {\it the absence of spinon deconfinement
without a Berry phase term}. The latter critical point is at
$\kappa^* = 0$, $y^*_m = 0$ and $\rho^* \not= 0$. Remarkably, we
find that this new fixed point remains stable against instanton
excitations. From this analysis one sees that although off
criticality the instantons are relevant everywhere, they become
irrelevant at the quantum critical point. This allows spinon
deconfinement, which is different from the conclusion by Senthil
et al., since this novel critical point does not coincide with
their one. We refer to this fixed point as the charge "neutral" XY
one. {\it The XY fixed point is, instead, identified with the
quantum critical point studied by Bernevig et al., thus showing
that the deconfinement of spinons takes place even without the
Berry phase}. As a result we find that the system is described by
the critical field theory in Eq. (13) near the quantum critical
point.

Recently, it was reported the result of Monte-Carlo
simulation\cite{Japan_group} supporting {\it the existence of
deconfined spinons at the quantum critical point of the $O(3)$
nonlinear $\sigma$ model in the absence of the contribution of
Berry phase}. They claimed that critical fluctuations of bosonic
spinons at the quantum critical point result in the nonlocal
action of the gauge field and this contribution causes the
deconfinement of spinons\cite{Japan_group}.

\begin{figure}
\includegraphics[width=6cm]{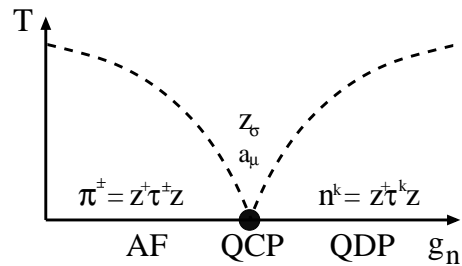}
\caption{\label{Fig. 1} Schematic phase diagram and proposed
elementary excitations in the O(3) nonlinear $\sigma$ model with
the spin stiffness $g_{n}^{-1}$}
\end{figure}

\section{Effective action for quantum antiferromagnets with easy
plane anisotropy: Abelian Higgs model with two flavors}

Low energy physics of two dimensional quantum antiferromagnets on
square lattices is described by the $O(3)$ nonlinear $\sigma$
model {\it in the presence of Berry
phase}\cite{Nagaosa_book,Berry_phase,Japan_group} \bqa && S =
S_{n} + S_{B} , \nn && S_{n} = \int{d^3x}
\frac{1}{2g_n}|\partial_{\mu}{\bf n}|^2 , \nn && S_{B} =
iS\sum_{r}\epsilon_{r}\mathcal{A}_{r} . \eqa Here ${\bf n}$ is the
unit three component vector representing the Neel order parameter.
$g_{n}^{-1}$ denotes the spin stiffness. The term $S_B$ represents
the contribution of Berry phase with $\epsilon_{r} = (-1)^{r_x +
r_y}$. $S$ in the Berry phase term is the value of spin $1/2$
here. $\mathcal{A}_{r}$ is the area enclosed by the curve mapped
out by the time evolution of ${\bf n}(\tau)$ on the unit
sphere\cite{Berry_phase}. Representing the spin component in terms
of bosonic spinons, ${\bf n} =
\frac{1}{2}z^{\dagger}_{\alpha}\sigma_{\alpha\beta}z_{\beta}$
called the $CP^1$ representation, we obtain an effective bosonic
quantum electrodynamics in two space and one time dimensions
($QED_3$)\cite{Nagaosa_book,Berry_phase,Japan_group} \bqa && S =
S_{B} + \int{d^3x} \Bigl[ \frac{1}{2g_n}|(\partial_{\mu} -
ia_{\mu})z_{\sigma}|^2 \Bigr] . \eqa Here $z_{\sigma}$ is the
bosonic spinon with $\sigma = 1, 2$ and $a_{\mu}$, the compact
U(1) gauge field mediating long range interactions among spinons.
$S_B$ is the Berry phase action in association with the time
component of the U(1) gauge field\cite{Berry_phase}. Following
Senthil et al., we consider easy plane anisotropy. In the easy
plane limit the bosonic spinor is represented to be $z_{\sigma} =
\left( \begin{array}{c} z_{1} \\ z_{2} \end{array} \right) =
\frac{1}{\sqrt{2}}\left( \begin{array}{c} e^{i\phi_1} \\
e^{i\phi_2} \end{array} \right)$\cite{Berry_phase}. Inserting this
into the above action Eq. (2), we obtain an effective field theory
for the SU(2) quantum antiferromagnet with the easy plane
anisotropy, {\it $N_{b} = 2$ Abelian Higgs model} in the field
theoretic language \bqa && S = \int{d^3x} \Bigl[
\frac{\rho}{2}|\partial_{\mu}\phi_1 - a_{\mu}|^2 +
\frac{\rho}{2}|\partial_{\mu}\phi_2 - a_{\mu}|^2 \nn && +
\frac{1}{2e^2}|\partial\times{a}|^2 \Bigr] .  \eqa Here $N_{b}$ is
the flavor number of bosonic spinons. As mentioned earlier, the
flavors are two ($N_b = 2$). $\rho\sim{g}_{n}^{-1}$ is the
stiffness parameter of the phase fields of spinons. The kinetic
energy of the gauge field is introduced with an internal gauge
charge $e$. The kinetic energy can be generated by integration
over high energy spinons. In $(2+1)D$ this term does not affect
the phase transitions of this model. This is because ${1}/{e^2}$
has a negative scaling dimension and this kinetic energy term
becomes irrelevant in the low energy limit. It is noted again that
the Berry phase term $S_B$ will not be considered any more. {\it
The present paper investigates the deconfinement of spinons at the
quantum critical point in the absence of the Berry phase effect}.

\section{Abelian Higgs model with one flavor:
relevance of instanton excitations at the IXY fixed point}

We first review the results of Senthil et al.\cite{Berry_phase}.
Although the $N_b = 1$ Abelian Higgs model is considered in this
section, this consideration shows well how Berry phase plays a
special role, causing the deconfinement of spinons. We note that
this one flavor Abelian Higgs model was also utilized to show the
relevance of Berry phase as a toy model in Ref.
\cite{Berry_phase}. We consider the following $N_b = 1$ Abelian
Higgs model \bqa && S = \int{d^3x} \Bigl[
\frac{\rho}{2}|\partial_{\mu}\phi - a_{\mu}|^2 +
\frac{1}{2e^2}|\partial\times{a}|^2 \Bigr]. \eqa Here $\phi$ is
the phase of a Higgs field and $a_\mu$, the compact U(1) gauge
field. $\rho$ is the phase stiffness parameter and $e$, the
internal electric charge of the Higgs field. This effective action
is usually proposed to describe a superconductor to insulator
transition of charged bosons\cite{Kleinert}. In this paper we
focus our attention on phase fluctuations instead of amplitude
fluctuations of Higgs fields. In the case of noncompact U(1) gauge
fields a charged fixed point to govern the superconducting
transition is expected to exist\cite{Kleinert}. $RG$ equations are
obtained to be in one loop level \bqa && \frac{d\rho}{dl} =
(D-2)\rho - \gamma{e^2}\rho , \nn && \frac{de^2}{dl} = (4-D)e^2 -
\lambda{e}^4 \eqa with $\gamma = \frac{2}{3\pi}$ and $\lambda =
\frac{1}{24\pi} $\cite{Wave_function_renormalization}. $l$ is a
usual scaling parameter and $D$ denotes a dimension of space and
time. We consider the case of $D =3$. The last term $-
\gamma{e^2}\rho$ in the first equation originates from the
self-energy correction of the Higgs field owing to gauge
fluctuations while the term $- \lambda{e}^4$ in the second
equation results from that of the gauge field due to screening of
the internal gauge charge by massless excitations of the Higgs
fields\cite{Wave_function_renormalization}. In Fig. 2 these
processes are explicitly shown by Feynman diagrams. In these $RG$
equations there exist two fixed points; one is the XY (neutral)
fixed point of $e^{*2} = 0$ and $\rho^{*} = 0$ and the other, the
IXY (charged) fixed point of $e^{*2} = \frac{1}{\lambda}$ and
$\rho^{*} = 0$. The XY fixed point is unstable against nonzero
charge $e^2 \not= 0$ and the $RG$ flows in the parameter space of
$(\rho,e^{2})$ converge into the IXY fixed point owing to $1 -
\gamma{e^{*2}} = 1 - \frac{\gamma}{\lambda} < 0$. In other words,
the quantum critical point of the superconductor to insulator
transition is described by the IXY fixed point\cite{Kleinert}.

\begin{figure}
\includegraphics[width=6cm]{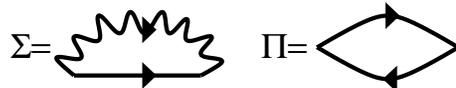}
\caption{\label{Fig. 2} Feynman diagrams of the self-energy of
Higgs fields and that of U(1) gauge fields}
\end{figure}

In the case of compact U(1) gauge fields we must admit instanton
excitations representing tunnelling events between topologically
inequivalent gauge vacua. Performing the standard duality
transformation\cite{Berry_phase,Kim_PRB,NaLee}, we obtain an
effective vortex action in the presence of instanton contributions
\bqa && S_{dual} = \int{d^3x} \Bigl[
\frac{\kappa}{2}|\partial_{\mu}\theta - c_{\mu}|^2 +
\frac{1}{2\rho}|\partial\times{c}|^2 \nn && +
\frac{e^2}{2}c_{\mu}^{2} - y_{m}cos\theta \Bigr] . \eqa Here
$\theta$ is the phase of a vortex field and $c_\mu$, the vortex
gauge field mediating interactions between the vortices. $\kappa$
is the stiffness parameter of the vortex phase field $\theta$ and
$y_m \sim e^{-S_{inst}}$, the instanton fugacity with an instanton
action $S_{inst} \sim {1}/{e^2}$\cite{NaLee}. {\it The vortex
gauge field $c_{\mu}$ is massive owing to the massless U(1) gauge
field $a_{\mu}$ and it can be ignored in the low energy
limit}\cite{Sudbo}. The last term $-y_{m}cos\theta $ appears as a
result of instanton excitations\cite{Berry_phase,NaLee}. When an
instanton is created with a probability $y_m$, a magnetic flux
should be emitted from the instanton owing to the gauss law. In
the presence of Higgs fields the magnetic flux is in the form of a
vortex. Thus a vortex creation operator $e^{-i\theta}$ is attached
to an instanton in the form of $y_{m}e^{-i\theta}$. Performing the
summation of instanton and anti-instanton excitations in the
dilute approximation, the $\cos$ potential for vortex fluctuations
is obtained\cite{Berry_phase,NaLee}. In the above sine-Gordon
action phase fluctuations of the vortex fields act as instanton
(magnetic) potentials to the instantons (Dirac magnetic
monopoles). Integrating over vortex phase fluctuations instead of
performing the summation of instantons, we obtain Coulomb
interactions $\sim 1/x$ in $(2+1)D$ between the
instantons\cite{Polyakov,NaLee}. Interaction strength of the
magnetic potential is proportional to $\kappa^{-1}$. Thus, the
inverse stiffness parameter $\kappa^{-1}$ plays the same role as
the magnetic charge. Owing to the Coulomb interaction the
instantons are expected to be deconfined. This implies that
tunnelling events are very activated. Gauge fluctuations $a_{\mu}$
are very strong and confinement of Higgs fields is obtained. In
the following we shall see this using an $RG$ analysis.

{\it Ignoring the vortex gauge field $c_\mu$}, we obtain the $RG$
equations of the usual sine-Gordon model\cite{Large_N_limit,KT}
\bqa && \frac{d\kappa}{dl} = (D-2)\kappa + {\beta}{y_m^2}
\frac{1}{\kappa}, \nn && \frac{dy_m}{dl} = (D -
\alpha\frac{1}{\kappa})y_m \eqa with positive numerical constants,
$\beta$ and $\alpha$\cite{Large_N_limit}. In our consideration
their precise values are not important. In these two equations
there exist no stable fixed points in $(2+1)D$ while in $(1+1)D$
there is a line of fixed points describing the Kosterliz-Thouless
transition as well known\cite{KT}. The fixed point of $\kappa^*
=0$ and $y_m^* = 0$ corresponds to the IXY fixed point of the
original Abelian Higgs model. The IXY fixed point is not stable
against instanton excitations $y_{m} \not= 0$. Both the phase
stiffness $\kappa$ and the instanton fugacity $y_m$ become larger
and larger at low energy. If we rewrite the $RG$ equation of the
stiffness parameter $\kappa$ in terms of the magnetic charge $g$
corresponding to the inverse stiffness parameter $\kappa^{-1}$ in
the presence of Higgs fields, i.e., $g = \kappa^{-1}$, we obtain
the same $RG$ equation with Ref. \cite{Large_N_limit} for the
magnetic charge, $dg/dl = -(D-2)g - \beta{y}_{m}^{3}g^{3}$. The
effective magnetic charge $g$ becomes smaller and smaller to be
zero owing to the negative bare scaling dimension $-(D-2)$ in the
presence of screening of magnetic charges by instanton
excitations. The negative bare scaling dimension $-1$ of the
magnetic charge in $D = 3$ results from the bare Coulomb
interaction $\sim 1/x$ between instantons. The screening effect is
represented by the last term $- \beta{y}_{m}^{3}g^{3}$ and this
leads the Coulomb potential to be the Yukawa-type potential $\sim
e^{-x/\lambda}/x$ where $\lambda$ is the screening length in
association with the instanton fugacity. The zero magnetic charge
leads the instanton fugacity to go to infinity at low energy. In
other words, the instantons are more activated. Depth of the
$\cos$ potential in Eq. (6) becomes deeper and deeper. Thus, the
phase of vortex fields is pinned at one ground position of the
$\cos$ potential. We conclude that instanton excitations induce
vortex condensation $<e^{i\theta}> \not= 0$ and instantons remain
deconfined as the case of the pure U(1) gauge theory in
$(2+1)D$\cite{Polyakov}. Confinement of charged bosons is
realized. This is the result in the absence of the Berry phase
effect.

If there exists a Berry phase term, the IXY fixed point can be
stable against instanton excitations\cite{Berry_phase}. The
contribution of Berry phase to instantons is given by ${\cal
L}_{B} =
i\frac{\pi}{2}\sum_{n}\zeta_{n}\Delta{Q}_{n}$\cite{Berry_phase,Japan_group}.
Here $n$ labels dual lattices of original lattices.
$\Delta{Q}_{n}$ represents an instanton excitation at the dual
site $n$. $\zeta_{n}$ is a fixed integer field and it is given by
$0, 1, 2, 3$ depending on whether the dual lattice coordinate is
(even, even), (even, odd), (odd, even) or (odd,
odd)\cite{Berry_phase,Japan_group}. {\it Performing the duality
transformation in the presence of this Berry phase
term}\cite{Berry_phase}, we find that the Berry phase gives rise
to spacial oscillation in the instanton induced term $-
y_{m}cos\theta$. Another way to say this is that the Berry phase
gives destructive interference to instanton excitations.
Instantons acquire Berry phases depending on instanton positions
and summation of the instantons results in spacial dependence in
the $\cos$ potential\cite{Berry_phase}. Thus, the contribution of
instanton excitations makes a partition function vanish unless the
instanton excitations are quadrupled, i.e., $\Delta{Q}_{n} \equiv
0$ ($mod$ $4$). {\it Only quadrupled instanton excitations
contribute to the partition function and this effect is proven to
be irrelevant at the quantum critical point described by the IXY
fixed point}\cite{Berry_phase,Dual_action}. It is the result of
Senthil et al. in the case of the $N_b = 1$ Abelian Higgs model.
In the case of the $N_b = 2$ Abelian Higgs model the same argument
can be applied to the IXY fixed point. As a result the
deconfinement of spinons is realized owing to the Berry phase
effect. However, we find that one different thing appears in the
case of the $N_b = 2$ Abelian Higgs model. We show that there
exists another fixed point called the charge neutral XY fixed
point. {\it The neutral XY fixed point is shown to be stable
against instanton excitations in the absence of Berry phase}.

\section{Renormalization group analysis of Abelian Higgs model with
two flavors: irrelevance of instanton excitations at the XY fixed
point}

We return to the main problem, $N_b = 2$ Abelian Higgs model Eq.
(3). In two spacial dimensions it is well known that the O(3)
nonlinear $\sigma$ model in the absence of Berry phase, $S_n$ in
Eq. (1) shows a continuous phase transition between an
antiferromagnetically ordered state with O(3) symmetry breaking
and a quantum disordered phase with no symmetry breaking at zero
temperature, depending on the spin stiffness parameter
$g_{n}^{-1}$\cite{Nagaosa_book}. Thus, Eq. (3) derived from Eq.
(1) is naturally expected to exhibit the second order quantum
phase transition between the two phases, depending on the phase
stiffness parameter $\rho$. At a sufficiently large stiffness
above the critical stiffness $\rho_{c}$ the Neel state would
emerge. This ordered phase is represented by condensation of
spinons, $<z_{\sigma}> \sim <e^{i\phi_{\sigma}}> \not= 0$. The
spinon condensation leads the U(1) gauge field to be massive via
the Anderson-Higgs mechanism. Integrating over the massive U(1)
gauge field, we obtain an effective field theory in terms of
spinon and anti-spinon confined objects, $z_{1}^{\dagger}z_{2}
\sim e^{-i(\phi_{1} - \phi_{2})}$ corresponding to
antiferromagnons of spin $1$. In the context of the gauge theory
this phase corresponds to the Higgs-confinement
phase\cite{Fradkin}. At a sufficiently small stiffness below the
critical stiffness quantum fluctuations of the phase fields
$\phi_{\sigma}$ destroy the antiferromagnetic long range order and
a quantum disordered phase appears with gapped spinons,
$<z_{\sigma}> \sim <e^{i\phi_{\sigma}}> = 0$. The massive spinon
excitations would be confined to form spinon and anti-spinon
composites corresponding to massive spin excitons or gapped
paramagnons of spin $1$.

It should be noted that the quantum disordered phase considered
above is fully symmetric and thus featureless {\it owing to the
absence of Berry phase}. If the Berry phase term is introduced in
the featureless quantum disordered phase, the Berry phase leads
the quantum disordered phase to be a valance bond solid
($VBS$)\cite{Berry_phase}. This $VBS$ exhibits translational
symmetry breaking\cite{Berry_phase}. In the $VBS$ massive spinons
are also confined to form massive meson excitations. These are
spin excitons, spin singlet to triplet
excitations\cite{Berry_phase}. In the context of the gauge theory
the $VBS$ corresponds to a confinement phase owing to the
condensation of instantons as the above quantum disordered phase.
In both the antiferromagnetism and the quantum disordered
paramagnetism {\it in the absence of Berry phase} or the $VBS$
{\it in the presence of Berry phase}, the spinons are always
confined and thus, fractional spin $1/2$ excitations, spinons are
not found.

Now we examine the quantum critical point. In the presence of the
Berry phase effect it was already discussed in the previous
section that the fractional spin $1/2$ spinon excitations can be
deconfined to emerge at the quantum critical point described by
the IXY fixed point. On the other hand, in the absence of Berry
phase the IXY fixed point was shown to be unstable against
instanton excitations. Only the confinement of spinons is expected
to arise. In this respect a new scenario for deconfinement of
spinons is necessary in the present case. It seems to be natural
to consider a new fixed point instead of the IXY fixed point. {\it
Indeed, a new stable fixed point, the charge neutral XY fixed
point is found in the $N_b = 2$ Abelian Higgs model}. The
deconfinement scenario in the present paper is completely
different from the previous one\cite{Berry_phase}.

Performing the standard duality transformation of the $N_b = 2$
Abelian Higgs model Eq. (3), we obtain an effective vortex action
in the presence of instantons\cite{Berry_phase,NaLee} \bqa &&
S_{dual} = \int{d^3x} \Bigl[ |(\partial_{\mu} -
ic_{1\mu})\Phi_1|^2 + |(\partial_{\mu} - ic_{2\mu})\Phi_2|^2 \nn
&& + m^2(|\Phi_1|^2 + |\Phi_2|^2) + \frac{u}{2}(|\Phi_1|^{4} +
|\Phi_2|^{4}) \nn && + \frac{1}{2\rho}|\partial\times{c_1}|^2 +
\frac{1}{2\rho}|\partial\times{c_2}|^2 + \frac{e^2}{2}|c_{1\mu} +
c_{2\mu}|^2 \nn && - z_{m}(\Phi_1^{\dagger}\Phi_2^{\dagger}+
\Phi_1\Phi_2 ) \Bigr] . \eqa Here $\Phi_{1(2)}$ represents the
vortex field and $c_{1(2)\mu}$, the vortex gauge field mediating
interactions between vortices. $m$ is the mass of vortex fields
and $u$, the coupling strength of local interactions between
vortices. The vortex mass is given by $m^2 \sim \rho - \rho_{c}$,
where $\rho_{c}$ is the critical stiffness parameter. $z_m \sim
e^{-S_{inst}}$ is the instanton fugacity with an instanton action
$S_{inst} \sim {1}/{e^2}$. Vortex excitations can be considered to
be merons (half skyrmions)\cite{Berry_phase}. $\Phi_{1}$ is the
vortex in the $z_{1} \sim e^{i\phi_{1}}$ spinon field and it
carries down spin ${\bf n}^{z} = - 1/2$ in the
core\cite{Meron_Spin}. $\Phi_{2}$ is that in the $z_{2} \sim
e^{i\phi_{2}}$ spinon field and it carries up spin ${\bf n}^{z} =
1/2$ in the core\cite{Meron_Spin}. Physical picture of the meron
excitations is well described in Ref. \cite{Berry_phase}. When an
instanton is created with a probability $z_m$, a magnetic flux
should be emitted from the instanton owing to the gauss law. Since
there exist two kinds of vortices, the vortex creation operator
$\Phi^{\dagger}_{1}\Phi^{\dagger}_{2}$ is attached to the
instanton\cite{Berry_phase,NaLee}. Here we should not forget the
spin degrees of freedom in the meron fields. Then, we can see that
the operator $\Phi^{\dagger}_{1}\Phi^{\dagger}_{2}$ represents a
skyrmion excitation\cite{Definition}. The spin up meron $\Phi_{2}$
turns into the spin down meron $\Phi_{1}$ and vice
versa\cite{Definition}. In this respect an instanton excitation
represents a tunnelling event between the spin down and spin up
merons, corresponding to a skyrmion (hedgehog) configuration of
the Neel vector fields, ${\bf n}$. In this dual vortex formulation
it is the main problem whether the instanton induced term
representing skyrmion excitations is relevant or not. If this term
is relevant in the $RG$ sense, only skyrmion excitations can
appear. The skyrmion excitations change spin $1$ (${\bf
n}^{z}_{2}$ $-$ ${\bf n}^{z}_{1} = 1$) in the vortex core. As a
result only spin $1$ excitations are possible and fractionalized
spinon excitations do not occur. In other words, a spin up meron
is confined with a spin down meron to appear only in the form of a
skyrmion. Only if the skyrmion excitation term becomes irrelevant,
the meron excitations of spin $1/2$ can emerge. The confinement
(deconfinement) of merons in the dual vortex description
corresponds to the confinement (deconfinement) of spinons in the
original Higgs field representation. It is known that the
instanton induced term is relevant in both antiferromagnetism and
quantum disordered paramagnetism\cite{Berry_phase}. We study the
relevance of the instanton induced term using an $RG$ analysis. In
the case of the $N_b = 1$ Abelian Higgs model we have already seen
the relevance of instanton excitations.

In passing, we briefly discuss vortex descriptions for possible
quantum phases. In the case of $\rho > \rho_{c}$ a vortex vacuum
$<\Phi_{\sigma}> = 0$ is energetically favorable. This corresponds
to antiferromagnetism where spinons are condensed, $<z_{\sigma}>
\not= 0$. As mentioned above, the skyrmion excitation term is
relevant and thus $<\Phi_{1}\Phi_{2}> \not= 0$ is obtained. In the
opposite case vortex condensation $<\Phi_{\sigma}> \not= 0$ is
expected to occur. This naturally leads to $<\Phi_{1}\Phi_{2}>
\not= 0$. The vortex condensation results in quantum disordered
paramagnetism where spinons are gapped, $<z_{\sigma}> = 0$ and
confined. The quantum critical point emerges at $\rho = \rho_{c}$.
In the following we show that instanton excitations become
irrelevant at the quantum critical point, thus causing
$<\Phi_{1}\Phi_{2}> = 0$. This implies deconfinement of meron
excitations, $\Phi_{1}$ and $\Phi_{2}$, corresponding to that of
spinon excitations, $z_{1}$ and $z_{2}$.

In Eq. (8) the mass term $\frac{e^2}{2}|c_{1\mu} + c_{2\mu}|^2$
resulting from the massless U(1) gauge field $a_{\mu}$ permits us
to set $c_{2\mu} = - c_{1\mu}\equiv - c_{\mu}$ in the low energy
limit. As a result we obtain the following dual vortex action in
the low energy limit \bqa && S_{dual} = \int{d^3x} \Bigl[
\frac{\kappa}{2}|\partial_{\mu}\theta_1 - c_{\mu}|^2 +
\frac{\kappa}{2}|\partial_{\mu}\theta_2 + c_{\mu}|^2 \nn && +
\frac{1}{2\rho}|\partial\times{c}|^2 - y_{m}cos(\theta_1+\theta_2)
\Bigr] .  \eqa Here $\theta_{1(2)}$ is the phase field of the
vortex field $\Phi_{1(2)}$. $\kappa$ is the stiffness parameter of
the vortex phase fields and $y_m =
2\bar{\Phi}_{1}\bar{\Phi}_{2}z_m$, the renormalized instanton
fugacity with the amplitude of vortex condensation
$\bar{\Phi}_{1(2)} = |<\Phi_{1(2)}>|$. We replaced ${\rho}/{2}$
with $\rho$. {\it In the above dual action one massless vortex
gauge field $c_\mu$ appears in contrast to the case of $N_b=1$,
Eq. (6) where there is no massless vortex gauge field}. We note
that in the vortex vacuum the massless vortex gauge fields
correspond to magnon excitations in the antiferromagnetic long
range order. In the following we show that {\it existence of the
massless vortex gauge field causes the instanton fugacity $y_m$ to
be zero at the quantum critical point even in the absence of Berry
phase}.

We first discuss two limiting cases in Eq. (9); one is $\rho
\rightarrow 0$ which allows us to ignore the vortex gauge field
and the other, $y_m \rightarrow 0$ which permits us to ignore the
instanton excitations. First, ignoring the vortex gauge field in
Eq. (9), we obtain the following RG equations \bqa &&
\frac{d\kappa}{dl} = (D-2)\kappa + {\beta}{y_m^2}
\frac{2}{\kappa}, \nn && \frac{dy_m}{dl} = (D -
\alpha\frac{2}{\kappa})y_m . \eqa In the case when the vortex
gauge field is ignored, the vortex Lagrangian Eq. (9) is the same
as the Lagrangian Eq. (6) except the fact that the flavor number
is two in Eq. (9). The effective magnetic charge, $g =
\kappa^{-1}$ is screened by two kinds of vortices. If we rewrite
the first $RG$ equation in Eq. (10) in terms of the effective
magnetic charge $g$, we obtain $\frac{dg}{dl} = - (D-2)g -
2\beta{y}_{m}^{2}g^{3}$. As shown by the second term, two kinds of
vortices screen out the magnetic charge. Eq. (10) is the same as
Eq. (7) except the factor $2$. In an appendix we briefly sketch
how Eq. (10) is derived from Eq. (9) in the absence of the vortex
gauge field $c_{\mu}$. In these $RG$ equations both $\kappa$ and
$y_m$ become larger and larger in the low energy limit as the case
of the $N_b =1$ Abelian Higgs model [Eq. (7)]. There exist no
stable fixed points. Instanton excitations are relevant and only
the confinement of meron fields $\theta_{1(2)}$ (the confinement
of spinon fields $\phi_{1(2)}$) is expected to occur. Next,
ignoring the instanton excitations, i.e., the compactness of the
U(1) gauge field $a_{\mu}$ in Eq. (9), we obtain the same form of
Lagrangian as Eq. (4) and get similar $RG$ equations with Eq.
(5)\cite{Wave_function_renormalization} \bqa && \frac{d\kappa}{dl}
= (D-2)\kappa - \gamma{\rho}{\kappa}, \nn && \frac{d\rho}{dl} =
(4-D)\rho - 2\lambda{\rho^2} . \eqa The factor $2$ in the second
equation results from the screening effect by two flavors of the
vortex fields. In the above we have two fixed points; one is the
IXY fixed point of $\rho^* = 0$ and $\kappa^* = 0$ which is
unstable against nonzero value of $\rho$ and the other, the stable
XY fixed point of $\rho^* = \frac{1}{2\lambda}$ and $\kappa^* =
0$\cite{XY_fixed_point}. The stability is guaranteed by
$1-\gamma\rho^{*} = 1 - \frac{\gamma}{2\lambda} < 0$. Consider one
to one correspondence of the $RG$ equations between Eq. (5) and
Eq. (11). Note that if we ignore the instanton excitations in Eq.
(7), we obtain the fixed point of $\kappa^* = 0$. This fixed point
of the dual vortex action Eq. (6) in the absence of instantons
corresponds to the IXY fixed point of a superconductor to
insulator transition in the original Higgs field representation
Eq. (4). What Eq. (7) and Eq. (10) tell us is that the IXY fixed
point becomes unstable when we admit the instanton excitations.
{\it The presence of the additional vortex gauge field also makes
the IXY fixed point unstable even in the absence of instanton
excitations, resulting in the stable XY fixed point in the case of
the same stiffness parameter for the two phase fields. It is the
key question in this paper whether the XY fixed point in Eq. (11)
remains stable or not after including the instanton excitations}.

Admitting both the massless vortex gauge fields and the instanton
excitations, we obtain the following $RG$ equations {\it as a
combined form of Eq. (10) and Eq. (11)} \bqa && \frac{d\kappa}{dl}
= (D-2)\kappa + {\beta}{y_m^2} \frac{2}{\kappa}  -
\gamma{\rho}{\kappa}, \nn && \frac{d\rho}{dl} = (4-D)\rho -
2\lambda{\rho^2} , \nn && \frac{dy_m}{dl} = (D -
\alpha\frac{2}{\kappa})y_m . \eqa In these $RG$ equations the XY
fixed point of $\rho^* = \frac{1}{2\lambda}$, $\kappa^* = 0$ and
$y_{m}^{*} = 0$ is only the stable one against instanton
excitations while the IXY fixed point of $\rho^* = 0$, $\kappa^* =
0$ and $y_{m}^{*} = 0$ is unstable against both the vortex gauge
field excitations $\rho \not= 0$ and the instanton excitations
$y_{m} \not= 0$. It is instructive to rewrite the first $RG$
equation in terms of the effective magnetic charge $g =
\kappa^{-1}$. It is obtained to be $\frac{dg}{dl} = -(D-2)g -
2{\beta}{y_m^2}g^{3} + \gamma{\rho}g$. {\it At the XY fixed point
the effective magnetic charge $g^{*} = \kappa^{*-1}$ becomes
infinite because of $- (1-\gamma\rho^{*}) > 0$ as the case of
noncompact gauge fields}. Since the XY fixed point is the charge
neutral fixed point, it seems to be natural that the effective
magnetic charge in the presence of Higgs fields is infinite at the
XY fixed point\cite{EM_duality}. This infinitely large effective
magnetic charge makes the instanton excitations irrelevant, i.e.,
$y_m \rightarrow 0$.

In both the Neel and quantum disordered phases the instanton
induced term plays a special role, resulting in only the skyrmion
excitations. Away from the quantum critical point (the XY fixed
point) we find that the instanton fugacity goes to infinity in the
low energy limit [Eq. (12)]. Thus, depth of the $\cos$ potential
in Eq. (9) becomes infinitely deep in the low energy limit and one
ground position is chosen for the $\theta_{1} + \theta_{2}$ field.
This implies that fluctuations of the $\theta_{1}$ field are
strongly correlated with those of the $\theta_2$ field,
permanently causing an only ground state of the $\cos$ potential
for the $\theta_{1} + \theta_{2}$ field. This leads to
$<\Phi_{1}\Phi_{2}> \not= 0$ in both phases of Eq. (8). Thus, the
meron excitations are not possible and only spin $1$ excitations
are expected to occur\cite{Berry_phase,NaLee}. However, at the
quantum critical point the instanton excitations become irrelevant
as shown in Eq. (12). The $\cos$ potential in Eq. (9) can be
safely ignored at the quantum critical point and the $\theta_{1}$
field can fluctuate "independently" with the $\theta_2$ field.
Here "" is used in the sense that the $\theta_1$ field is coupled
to the $\theta_2$ field via the noncompact U(1) gauge field
$c_{\mu}$. As a result the meron excitations carrying
fractionalized spin ${1}/{2}$ are expected to appear.

Now we can reach the critical field theory at the XY fixed point
based on the results of Eq. (12). Inserting the fixed point values
of $\rho^* = \frac{1}{2\lambda}$, $\kappa^* = 0$ and $y_{m}^{*} =
0$ into Eq. (9), we obtain the critical field theory at the XY
fixed point, ${\cal L}_{dual} =
\frac{\kappa^*}{2}|\partial_{\mu}\theta_1 - c_{\mu}|^2 +
\frac{\kappa^*}{2}|\partial_{\mu}\theta_2 + c_{\mu}|^2 +
\frac{1}{2\rho^*}|\partial\times{c}|^2 -
y^*_{m}cos(\theta_1+\theta_2) =
\frac{1}{2\rho^*}|\partial\times{c}|^2$. However, this critical
field theory is not satisfactory in the sense that there are no
terms representing critical fluctuations of vortex fields
(merons). {\it There should be deconfined critical meron
fluctuations}. It seems to be natural to introduce the
contribution of critical vortex fluctuations coupled to the
noncompact vortex gauge fields $c_{\mu}$, ${\cal L}_{v} =
|(\partial_{\mu} - ic_{\mu})\Phi_{1}|^2 + |(\partial_{\mu} +
ic_{\mu})\Phi_{2}|^2 + m^{*2}(|\Phi_1|^2 + |\Phi_2|^2) +
\frac{u^{*}}{2}(|\Phi_1|^{4} + |\Phi_2|^{4})$. Here $\Phi_{1(2)}$
represents the meron field. $m^{*}$ and $u^{*}$ are the fixed
point values of the mass and self-interaction strength of vortex
fields, respectively, at the quantum critical point. Notice that
the fixed point value of the vortex mass should be zero, $m^{*2} =
0$. This zero vortex mass trivially leads to the zero fixed point
value of the vortex stiffness parameter, i.e., $\kappa^* = 0$ at
the quantum critical point. This can be easily checked by the
relation $\kappa^* \sim - m^{*2}/u^* = 0$ in the mean field (tree)
level. The fixed point value $u^*$ is not explicitly shown in the
present paper since we utilize the effective phase action Eq. (9).
It is certain that its fixed point value is finite\cite{Kleinert}.
As a result we reach the following critical field theory \bqa &&
{\cal L}_{c} = |(\partial_{\mu} - ic_{\mu})\Phi_{1}|^2 +
|(\partial_{\mu} + ic_{\mu})\Phi_{2}|^2 \nn && +
\frac{u^{*}}{2}(|\Phi_{1}|^{4} + |\Phi_{2}|^{4}) +
\frac{1}{2\rho^{*}}|\partial\times{c}|^2 . \eqa {\it The
deconfined quantum critical point in the O(3) nonlinear $\sigma$
model with the easy plane anisotropy is described by the critical
field theory Eq. (13) in terms of the merons interacting via the
noncompact U(1) gauge fields}.

We reemphasize the main difference between our deconfinement
scenario and the previous one\cite{Berry_phase}. In the earlier
study\cite{Berry_phase} the fixed point to govern critical
dynamics was not clearly pointed out. But, the
study\cite{Berry_phase} infers that the fixed point is the charged
fixed point (IXY fixed point). {\it At the charged fixed point the
Berry phase plays a special role (causing destructive interference
for instanton excitations and making the instantons irrelevant) to
result in the deconfinement of critical bosonic spinons}. On the
other hand, our present study claims that the true fixed point of
the quantum phase transition in the O(3) nonlinear $\sigma$ model
with the easy plane anisotropy is not the charged fixed point but
the charge neutral fixed point (XY fixed point). {\it At the
neutral XY fixed point the fixed point value of the internal
charge is zero and thus its corresponding magnetic charge is
infinite, causing the irrelevance of instantons even in the
absence of Berry phase}. The deconfinement of spinons is expected
to occur at the quantum critical point.

One may suspect that it is physically meaningful to consider the
O(3) nonlinear $\sigma$ model without Berry phase. In a different
angle this doubt is associated with the question {\it when the
contribution of Berry phase can be ignored}. The following two
cases may be the candidates; one is the case of {\it double
layered antiferromagnets} and the other, {\it the presence of
disorders}. It is well known that in two leg ladders the
contribution of Berry phase cancels between the legs\cite{Ladder}.
The same mechanism works in the double layered quantum
antiferromagnets\cite{Chubukov}. In this case the mechanism of
spinon deconfinement proposed by Senthil et al. cannot be applied.
Instead, our mechanism may be applicable. One problem is that in
the double layered antiferromagnet there exist more flavors than
those in the one layer system. However, it is certain that
massless vortex gauge fields still remain. The presence of
massless vortex gauge fields is expected to cause the
deconfinement of spinons. More cautious studies are required near
future.

The presence of nonmagnetic disorders leads to random depletion of
spins. This results in two important effects. First, the random
depletion introduces a random Berry phase term to the nonlinear
$\sigma$ model\cite{RandomBerry}. Second, it causes a random
exchange coupling between spins\cite{RandomBerry}. We expect that
the random Berry phase term is difficult to suppress instanton
excitations. The contribution of Berry phase to instantons is
given by ${\cal L}_{B} =
i\frac{\pi}{2}\sum_{n}\zeta_{n}\Delta{Q}_{n}$\cite{Berry_phase},
as mentioned earlier. Remember that in the absence of randomness
$\zeta_{n}$ is a fixed integer field and it is given by $0, 1, 2,
3$ depending on whether the dual lattice coordinate is (even,
even), (even, odd), (odd, even) or (odd, odd)\cite{Berry_phase}.
{\it The presence of disorders introduces randomness to
$\zeta_{n}$}. In other words, the random depletion of spins
results in $<\zeta_{n}> = 0$, where $<...>$ denotes the average
over disorders. The effect of random Berry phase would not be
sufficient to suppress the instanton excitations. Thus, the
mechanism of spinon deconfinement by Senthil et al. would not work
in the presence of disorders. One problem is the effect of random
exchange couplings. In the limit of weak randomness we may treat
the effect of disorders as a random mass term of spinons in the
$CP^1$ representation\cite{Random}. Then, the problem becomes
whether the randomness is relevant or not. When the random mass is
relevant, the spinons would be localized near the disorders. The
quantum criticality is expected to disappear. As a result the
spinons would be confined to form antiferromagnetic spin
fluctuations of spin $1$. This consideration is consistent with
increase of antiferromagnetic correlations when nonmagnetic
impurities are doped into nonmagnetic states\cite{RandomBerry}.
Our preliminary calculation shows that the deconfined quantum
criticality is sustained against sufficiently weak disorders,
which is completely consistent with the case of fermionic $QED_3$
describing the algebraic spin liquid\cite{Kim_disorder}. In this
case the mechanism of the deconfined quantum criticality is due to
the XY fixed point resulting from massless vortex gauge fields.
The role of nonmagnetic disorders in the deconfined quantum
criticality is under our current investigation.

\section{Conclusion}

In the present study we investigated the deconfinement of bosonic
spinons at the quantum critical point of the O(3) nonlinear
$\sigma$ model without Berry phase in the easy plane limit. The
low energy effective field theory in the $CP^1$ representation is
given by the $N_b=2$ Abelian Higgs model with $N_b$, the flavor
number of bosonic spinons. {\it The quantum critical point of the
$N_b =2$ noncompact Abelian Higgs model corresponds to the XY
fixed point while that of the $N_b =1$ noncompact Abelian Higgs
model, the IXY fixed point}. {\it This difference originates from
the existence of massless vortex gauge fields in the case of $N_b
= 2$}. We showed that the instanton fugacity becomes zero at the
XY fixed point and thus, instanton excitations do not destabilize
the XY fixed point. As a consequence we find the critical field
theory [Eq. (13)] in terms of fractional particles (merons)
coupled to noncompact U(1) gauge fields at the quantum critical
point of the $N_b=2$ Abelian Higgs model in $(2+1)D$. On the other
hand, the IXY fixed point was shown to be unstable against
instanton excitations. In order to obtain deconfined spinons at
the IXY fixed point, the contribution of Berry phase seems to be
crucial.

\begin{acknowledgments}
We should acknowledge that most parts of the introduction are from
the comments of the fourth referee. We really appreciate his
contribution. We especially thank Dr. Yee, Ho-Ung for helpful
discussions of Eq. (5). We also thank prof. Kim, Yong-Baek for
pointing out that the study of Senthil et al. gives an existence
proof for the deconfinement of spinons.
\end{acknowledgments}

\appendix
\section{}

We briefly sketch how to derive Eq. (10) from Eq. (9) in the
absence of the vortex gauge field. This derivation is based on
Ref. \cite{Nagaosa_book}. We consider the two flavor sine-Gordon
action \bqa && S = \int{d^Dx} \Bigl[
\frac{\kappa}{2}(\partial_{\mu}\theta_{1})^2 +
\frac{\kappa}{2}(\partial_{\mu}\theta_{2})^2 -
y_{m}\cos(\theta_{1} + \theta_{2}) \Bigr] . \eqa Here we utilize
an Wilsonian approach. We first divide the
$\theta_{\sigma\Lambda}$ field defined on the momentum cut-off
$\Lambda$ into low and high energy degrees of freedom,
$\theta_{\sigma\Lambda'}$ and $h_{\sigma}$, respectively, \bqa &&
\theta_{\sigma\Lambda}(x) = \theta_{\sigma\Lambda'}(x) +
h_{\sigma}(x) , \nn && \theta_{\sigma\Lambda'}(x) =
\int_{0<p<\Lambda'}\frac{d^Dp}{(2\pi)^{D}}
e^{ip\cdot{x}}\theta_{\sigma\Lambda}(p) , \nn && h_{\sigma}(x) =
\int_{\Lambda'<p<\Lambda}\frac{d^Dp}{(2\pi)^{D}}
e^{ip\cdot{x}}\theta_{\sigma\Lambda}(p) . \eqa Inserting these low
and high energy degrees of freedom into the above Eq. (A1) and
integrating over the high energy field variables $h_{\sigma}$, we
obtain the following expression of a partition function \bqa &&
Z_{\Lambda} =
\int{D\theta_{1\Lambda}}{D\theta_{2\Lambda}}e^{-S_{\Lambda}[\theta_{1\Lambda},\theta_{2\Lambda}]}
\nn && = \int{D\theta_{1\Lambda'}Dh_{1}D\theta_{2\Lambda'}Dh_{2}}
e^{-S_{\Lambda}[\theta_{1\Lambda'}+h_{1},\theta_{2\Lambda'}+h_{2}]}
\nn && = \int{D\theta_{1\Lambda'}D\theta_{2\Lambda'}} e^{-{\tilde
S}_{\Lambda}[\theta_{1\Lambda'},\theta_{2\Lambda'}]} . \eqa Here
the effective action ${\tilde
S}_{\Lambda}[\theta_{1\Lambda'},\theta_{2\Lambda'}]$ defined on
the momentum cut-off $\Lambda$ is given by   \bqa && e^{-{\tilde
S}_{\Lambda}[\theta_{1\Lambda'},\theta_{2\Lambda'}]} \nn && =
\int{Dh_{1}Dh_{2}} exp\Bigl[ -\int{d^Dx} \Bigl(
\frac{\kappa}{2}(\partial_{\mu}\theta_{1\Lambda'})^2 +
\frac{\kappa}{2}(\partial_{\mu}\theta_{2\Lambda'})^2 \nn && +
\frac{\kappa}{2}(\partial_{\mu}h_{1})^2 +
\frac{\kappa}{2}(\partial_{\mu}h_{2})^2 -
y_{m}\cos(\theta_{1\Lambda'} + \theta_{2\Lambda'} + h_{1} + h_{2})
\Bigr) \Bigr] \nn && \equiv N{\exp}\Bigl[ -\int{d^Dx} \Bigl(
\frac{\kappa}{2}(\partial_{\mu}\theta_{1\Lambda'})^2 +
\frac{\kappa}{2}(\partial_{\mu}\theta_{2\Lambda'})^2 \Bigr)
\Bigr]\nn&&\Bigl<e^{\int{d^Dx} y_{m}\cos(\theta_{1\Lambda'} +
\theta_{2\Lambda'} + h_{1} + h_{2})} \Bigr>_{h_1,h_2} , \eqa where
$\Bigl< ... \Bigr>_{h_1,h_2}$ represents averaging over the
gaussian action of the high energy fields, $S_{h}[h_1,h_2] =
\int{d^Dx} \Bigl(\frac{\kappa}{2}(\partial_{\mu}h_{1})^2 +
\frac{\kappa}{2}(\partial_{\mu}h_{2})^2 \Bigr)$, and the constant
$N$ is given by $N = \int{Dh_1}{Dh_2}e^{-S_{h}[h_1,h_2]}$.
Expanding the exponential to the second order in the fugacity
($y_{m}$) expansion, we obtain the following expression of the
effective action ${\tilde
S}_{\Lambda}[\theta_{1\Lambda'},\theta_{2\Lambda'}]$, \bqa &&
{\tilde S}_{\Lambda}[\theta_{1\Lambda'},\theta_{2\Lambda'}] =
\int{d^Dx} \Bigl(
\frac{\kappa}{2}(\partial_{\mu}\theta_{1\Lambda'}(x))^2 +
\frac{\kappa}{2}(\partial_{\mu}\theta_{2\Lambda'}(x))^2 \nn && -
y_{m}\Bigl<\cos(\theta_{1\Lambda'}(x) + \theta_{2\Lambda'}(x) +
h_{1}(x) + h_{2}(x))\Bigr>_{h_1,h_2} \Bigr) \nn && -
\int{d^Dx}\int{d^Dx'} \frac{y_{m}^2}{2}\Bigl(
\Bigl<cos(\theta_{1\Lambda'}(x) + \theta_{2\Lambda'}(x) \nn && +
h_{1}(x) + h_{2}(x)) \cos(\theta_{1\Lambda'}(x') +
\theta_{2\Lambda'}(x') \nn && + h_{1}(x') +
h_{2}(x'))\Bigr>_{h_1,h_2} - \Bigl<\cos(\theta_{1\Lambda'}(x) +
\theta_{2\Lambda'}(x) \nn && + h_{1}(x) +
h_{2}(x))\Bigr>_{h_1,h_2} \Bigl<\cos(\theta_{1\Lambda'}(x') +
\theta_{2\Lambda'}(x') \nn && + h_{1}(x') +
h_{2}(x'))\Bigr>_{h_1,h_2} \Bigr) . \eqa

Now we evaluate the average of the $\cos$ potentials over the
gaussian action $S_{h}[h_1,h_2]$ of the high energy fields $h_1,
h_2$. The term of the first order in the fugacity $y_m$ is
obtained to be \bqa && \Bigl<\cos(\theta_{1\Lambda'}(x) +
\theta_{2\Lambda'}(x) + h_{1}(x) + h_{2}(x))\Bigr>_{h_1,h_2} \nn
&& = \frac{1}{2}\Bigl(e^{i\theta_{1\Lambda'}(x) +
i\theta_{2\Lambda'}(x)}\Bigl<e^{ih_{1}(x)+ih_2(x)}\Bigr>_{h_1,h_2}
+ h.c. \Bigr) \nn && = exp\Bigl[
-\frac{1}{2}G_{h_1}(0)-\frac{1}{2}G_{h_2}(0)\Bigr]\cos(\theta_{1\Lambda'}(x)
+ \theta_{2\Lambda'}(x)) \nn && =
B_1(0)B_2(0)\cos(\theta_{1\Lambda'}(x) + \theta_{2\Lambda'}(x)) .
\eqa Here $G_{h_\sigma}(x)$ is the propagator of the high energy
fields, given by $G_{h_\sigma}(x) =\frac{1}{\kappa}
\int_{\Lambda'<p<\Lambda}\frac{d^Dp}{(2\pi)^{D}}e^{ip\cdot{x}}\frac{1}{p^2}$,
and its associated factor $B_{\sigma}$, $B_{\sigma} =
exp[-\frac{1}{2}G_{h_\sigma}(0)]$. $G_{h_1} = G_{h_2}$ is
trivially shown, resulting in $B_{1} = B_{2}$. This is the reason
why the factor $2$ appears in Eq. (10). Notice from the momentum
integral that the quantities, $G_{\sigma}$ and $B_{\sigma}$ depend
on the momentum cut-off. The terms of the second order in the
fugacity $y_m$ can be calculated in the same way \bqa &&
\Bigl<\cos(\theta_{1\Lambda'}(x) + \theta_{2\Lambda'}(x) +
h_{1}(x) + h_{2}(x))\nn&&\cos(\theta_{1\Lambda'}(x') +
\theta_{2\Lambda'}(x') + h_{1}(x') + h_{2}(x'))\Bigr>_{h_1,h_2}
\nn && - \Bigl<\cos(\theta_{1\Lambda'}(x) + \theta_{2\Lambda'}(x)
+ h_{1}(x) + h_{2}(x))\Bigr>_{h_1,h_2} \nn&&\times
\Bigl<\cos(\theta_{1\Lambda'}(x') + \theta_{2\Lambda'}(x') +
h_{1}(x') + h_{2}(x'))\Bigr>_{h_1,h_2} \nn && =
\frac{1}{4}B_{1}^{2}(0)[B_{1}^2(x-x')-1]B_{2}^{2}(0)[B_{2}^2(x-x')-1]
\nn && \cos(\theta_{1\Lambda'}(x)+\theta_{1\Lambda'}(x') +
\theta_{2\Lambda'}(x)+\theta_{2\Lambda'}(x')) \nn && +
\frac{1}{4}B_{1}^2(0)[B_{1}^{-2}(x-x') -
1]B_{2}^2(0)[B_{2}^{-2}(x-x') - 1]\nn&&
\cos(\theta_{1\Lambda'}(x)-\theta_{1\Lambda'}(x')+
\theta_{2\Lambda'}(x)-\theta_{2\Lambda'}(x')) \nn && \approx
\frac{1}{4}B_{1}^2(0)[B_{1}^2(\xi) -1]B_{2}^2(0)[B_{2}^2(\xi)
-1]\nn&& \cos(2\theta_{1\Lambda'}(z)+2\theta_{2\Lambda'}(z)) \nn
&& + \frac{1}{4}B_{1}^2(0)[B_{1}^{-2}(\xi) -
1]B_{2}^2(0)[B_{2}^{-2}(\xi) - 1]\nn&&[1 -
\frac{1}{2}(\xi\cdot\partial\theta_{1\Lambda'}(z)+\xi\cdot\partial\theta_{2\Lambda'}(z))^2]
,\eqa where $z \equiv \frac{1}{2}(x+x')$ and $\xi \equiv x-x'$.

The last step in the Wilsonian $RG$ approach is the rescaling in
the coordinates $x$ and the momentum cut-off $\Lambda'$. Inserting
Eq. (A6) and Eq. (A7) into Eq. (A5), and performing the rescaling
$x \rightarrow e^{l}x'$ in the resulting effective action Eq. (5),
we obtain the following expression of the effective action \bqa &&
S_{\Lambda'}[\theta_{1\Lambda'},\theta_{2\Lambda'}] =
\int{d^Dx'}e^{Dl} \Bigl[ \nn && \frac{\kappa}{2}\Bigl(1
+\frac{y_{m}^2}{8\kappa}B_{1}^2(0)B_{2}^2(0)A
\Bigr)e^{-2l}(\partial_{\mu'}\theta_{1\Lambda'}(x'))^2 \nn && +
\frac{\kappa}{2}\Bigl(1
+\frac{y_{m}^2}{8\kappa}B_{1}^2(0)B_{2}^2(0)A
\Bigr)e^{-2l}(\partial_{\mu'}\theta_{2\Lambda'}(x'))^2 \nn && -
y_{m}B_{1}(0)B_{2}(0)\cos(\theta_{1\Lambda'}(x') +
\theta_{2\Lambda'}(x')) \Bigr] \nn && =\int{d^Dx'}\Bigl[
\frac{\kappa'}{2}(\partial_{\mu'}\theta_{1\Lambda'}(x'))^2 +
\frac{\kappa'}{2}(\partial_{\mu'}\theta_{2\Lambda'}(x'))^2 \nn &&
- y'_{m}\cos(\theta_{1\Lambda'}(x') + \theta_{2\Lambda'}(x'))
\Bigr]  \eqa with $A = \int{d^D\xi}[B_{1}^{-2}(\xi) -
1][B_{2}^{-2}(\xi) - 1]\xi^2$. As a result we find the scaling
relations between the renormalized and bare couplings \bqa &&
\kappa' = e^{(D-2)l}\Bigl(1
+\frac{y_{m}^2}{8\kappa}B_{1}^2(0)B_{2}^2(0)A \Bigr)\kappa , \nn
&& y'_{m} = e^{Dl}B_{1}(0)B_{2}(0)y_{m} . \eqa The above
expressions completely coincide with those in Ref.
\cite{Nagaosa_book} when the two flavors are reduced to one
flavor. Using the cut-off dependent green function
$G_{h_{\sigma}}(0) \sim \frac{1}{\kappa}l$ in $\Lambda' =
e^{-l}\Lambda$, we obtain the cut-off dependent values,
$B_{\sigma}(0) = e^{-\alpha\frac{1}{\kappa}l}$ and $A =
8\beta\frac{2}{\kappa}l$, where $\alpha$ and $\beta$ are positive
numerical constants. Inserting these into Eq. (A9) and expanding
the exponentials in the limit of $l \rightarrow 0$, we obtain the
$RG$ equations, Eq. (10) for the stiffness $\kappa$ and instanton
fugacity $y_{m}$. Notice that the two flavors $\sigma = 1, 2$ lead
to the numerical factor $2$ in Eq. (10).


\begin{thebibliography}{9}
\bibitem{Dual_action} In the argument of Ref. \cite{Berry_phase}
the dual vortex action in the presence of instanton excitations is
given by ${\cal L}_{dual} = |\partial_{\mu}\Phi_v|^{2} +
m_{v}^{2}|\Phi_v|^{2} + \frac{u_v}{2}|\Phi_{v}|^{4} -
y_{m}(\Phi_{v}^{n} + \Phi_{v}^{\dagger{n}})$. Here $\Phi_v$ is the
vortex field and $m_v$, $u_v$, its corresponding mass and
self-interaction strength, respectively. The last term is due to
instanton excitations, where $y_m$ is the instanton fugacity. When
there is no Berry phase contribution, $n$ is given by $1$. This
instanton induced effective potential is known to be relevant at
the critical point $m_{v}^{*} = 0$\cite{NaLee}. On the other hand,
the contribution of Berry phase results in $n = 4$ owing to
destructive interference of instanton events as discussed in the
text. This effective potential is irrelevant at the critical
point\cite{Berry_phase}.
\bibitem{Meron_Spin} The $z$ component spin is given by
$n^{z} = |<z_{1}>|^{2} - |<z_{2}>|^{2}$. The vortex $\Phi_{1}$
leads to $<z_{1}> = 0$ in the core and thus, it has down spin
$n^{z} = - |<z_{2}>|^{2} = - 1/2$. Owing to the similar reason the
vortex $\Phi_{2}$ has up spin $n^{z} = |<z_{1}>|^{2} = 1/2$ in the
core.
\bibitem{Definition} In the present paper the definition of $\Phi_{2}$
differs from that in Ref. \cite{Berry_phase}. If $\Phi_{2}$ is
transformed to $\Phi_{2}^{\dagger}$, Eq. (8) is rewritten to be
\bqa && S_{dual} = \int{d^3x} \Bigl[ |(\partial_{\mu} -
ic_{1\mu})\Phi_1|^2 + |(\partial_{\mu} + ic_{2\mu})\Phi_2|^2 \nn
&& + m^2(|\Phi_1|^2 + |\Phi_2|^2) + \frac{u}{2}(|\Phi_1|^{4} +
|\Phi_2|^{4}) \nn && + \frac{1}{2\rho}|\partial\times{c_1}|^2 +
\frac{1}{2\rho}|\partial\times{c_2}|^2 + \frac{e^2}{2}|c_{1\mu} +
c_{2\mu}|^2 \nn && - z_{m}(\Phi_1^{\dagger}\Phi_2+
\Phi_1\Phi_2^{\dagger} ) \Bigr] . \nonumber \eqa This dual vortex
action is the same as that in Ref. \cite{Berry_phase} in the
absence of Berry phase, if the vortex gauge field $c_{2\mu}$ is
replaced with $c_{2\mu} = - c_{1\mu} \equiv -c_{\mu}$. The
instanton induced terms describe tunnelling events associated with
spin flipping in vortex cores. In this respect instantons are
nothing but skyrmions.
\bibitem{EM_duality} Consider the electro-magnetic duality of
$eg = 1$, where $e$ is an internal electric charge and $g$, a
corresponding magnetic charge. If the electric charge becomes
small owing to screening by massless charged particles, the
magnetic charge gets large.
\bibitem{Random} Contributions of random exchange couplings may be approximated to
be random potentials or random masses of spinons in the $CP^1$
representation. The total exchange coupling $J_{ij}^{t}$ can be
considered to be $J_{ij}^{t} = J_{ij} + \delta{J}_{ij}$, where
$J_{ij}$ is the normal exchange coupling without randomness
between sites $i, j$ and $\delta{J}_{ij}$, the random exchange
coupling. The effective Lagrangian for the spinons is usually
obtained to be ${\cal L}_{z} = |(\partial_{\mu} -
ia_{\mu})z_{\sigma}|^2 + m_{rz}^{2}|z_{\sigma}|^{2} +
\frac{u_{z}}{2}(|z_{\sigma}|^{2})^{2}$ in the continuum
limit\cite{Berry_phase,Japan_group}. Here $m_{rz}$ is the total
mass of spinons and $u_z$, the strength of local interactions. The
total mass is expected to be $m_{rz}^{2} \sim J^{t} - J_{c} = J -
J_{c} + \delta{J} \equiv m_{z}^{2} + V$, where $J_c$ is the
critical coupling. The normal mass without disorders is given by
$m^{2}_{z} \sim J - J_{c}$. $V \sim \delta{J}$ can be considered
to be a random potential or random mass of spinons.
\end{thebibliography}
\end{document}